\documentclass[a4paper]{spie}  

\usepackage[]{graphicx}

\title{The GLAST Burst Monitor}

\author{Andreas von Kienlin\supit{a}, Charles A. Meegan\supit{b}, Giselher G. Lichti\supit{a}, 
Narayana P. Bhat\supit{c}, Michael S. Briggs\supit{c}, Valerie Connaughton\supit{c},  Roland Diehl\supit{a}, Gerald 
J. Fishman\supit{b}, Jochen Greiner\supit{a}, Andrew S. Hoover\supit{d}, R. Marc Kippen\supit{d}, Chryssa 
Kouveliotou\supit{e},   William S. Paciesas\supit{c}, Robert D. Preece\supit{a}, Volker Sch\"onfelder\supit{a}, 
Helmut Steinle\supit{a}, and Robert B. Wilson\supit{b}
\skiplinehalf
\supit{a}Max-Planck-Institut f\"ur extraterrestrische Physik, PO Box 1312, D-85741 Garching, Germany\\
\supit{b}NASA/Marshall Space-Flight Center, 320 Sparkman Drive, Huntsville, AL 35812, USA\\
\supit{c}University of Alabama, Huntsville, AL 35899, USA\\
\supit{d}Los Alamos National Laboratory, ISR-2, Mail Stop B244, Los Alamos, NM 87545, USA\\
\supit{e}Universities Space Research Association, USA
}

\authorinfo{Further author information: (Send correspondence to A. von Kienlin)\\A. v. Kienlin: E-mail: 
azk@mpe.mpg.de, Telephone: +49 (0)89 30000-3514\\  
C. A. Meegan: E-mail: charles.meegan@msfc.nasa.gov, Telephone:  (256) 961-7694 \\
G. G. Lichti: E-mail: grl@mpe.mpg.de, Telephone: +49 (0)89 30000-3536}

\begin{document} 
  \maketitle 

\begin{abstract}
The next large NASA mission in the field of gamma-ray astronomy, GLAST, is scheduled for launch in 2007. Aside from 
the main instrument LAT (Large-Area Telescope), a gamma-ray telescope for the energy range between 
20 MeV and $> 100$~GeV, a secondary instrument,  the GLAST burst monitor (GBM), is foreseen. With this monitor one of 
the key scientific objectives of the mission, the determination of  the high-energy behaviour of gamma-ray bursts and 
transients can be ensured. Its task is to increase the detection rate of gamma-ray bursts for the LAT and to extend 
the energy range to lower energies (from $\sim 10$ keV to $\sim 30$ MeV). It will  provide real-time burst locations 
over a wide FoV with sufficient accuracy to allow repointing the GLAST spacecraft.  Time-resolved spectra of many 
bursts recorded with LAT and the burst monitor will allow the investigation of the relation between the keV and the 
MeV-GeV emission from GRBs over unprecedented seven decades of energy. This will help to advance our understanding of 
the mechanisms by which gamma-rays are generated in gamma-ray bursts. 
\end{abstract}



\section{INTRODUCTION}
\label{sect:intro}  

GRBs are one of the most fascinating research topics in astrophysics. Since their discovery 35 years ago by the Vela 
satellites in 1967\cite{1973ApJ...182L..85K}, this phenomenon is still not totally understood and explained. The 
first major breakthrough in this field was obtained with the BATSE detectors\cite{1989GRO-SW..2...Fishman} on NASA's 
Compton Gamma Ray Observatory (CGRO) mission. In the nine years of the CGRO mission, about 3000 bursts were 
registered, which showed an isotropic distribution over the entire sky, but with a deficiency of weak bursts. It was 
the Italian/Dutch satellite  BeppoSAX\cite{1997A&AS..122..299B} which revealed the cosmological nature of GRBs with 
the identification of the first X-ray afterglow in 1997\cite{1997Natur.387..783C}, which triggered the first 
successful follow-up observation at optical wavelengths\cite{1997Natur.386..686V}.  This finally ruled out the 
Galactic population models. The redshifts obtained to date for about 36 GRBs range from 0.0085 to 
4.511\footnote{Table of GRBs with known redshifts: http://www.mpe.mpg.de/$\sim$jcg/grbrsh.html}. Observational 
evidence now strongly suggests that GRBs longer than 2 seconds are associated with 
hypernovae\cite{1998Natur.395..670G,2003ApJ...591L..17S,2003Natur.423..847H}.
The origin of shorter bursts is more of a mystery -- one hypothesis is that they are associated with the mergers of 
compact binaries.

The study of $\gamma$-ray bursts is one of the scientific objectives of the GLAST mission. This was motivated by 
observations of $\gamma$-ray bursts in the high-energy range above 50 MeV by EGRET onboard CGRO: The delayed emission 
of $\gamma$-quanta more than 1 hour after the burst start time was unexpected. In GRB940217 a 18 GeV $\gamma$-event 
was found in this extended emission \cite{1994Natur.372..652H}. This delayed emission is in contrast to the 
characteristics of most of the bursts observed in the BATSE energy range, which have only a maximum duration of 
several 100 sec. 

   \begin{figure}
   \begin{center}
   \begin{tabular}{c}
   \includegraphics[width=0.9\linewidth]{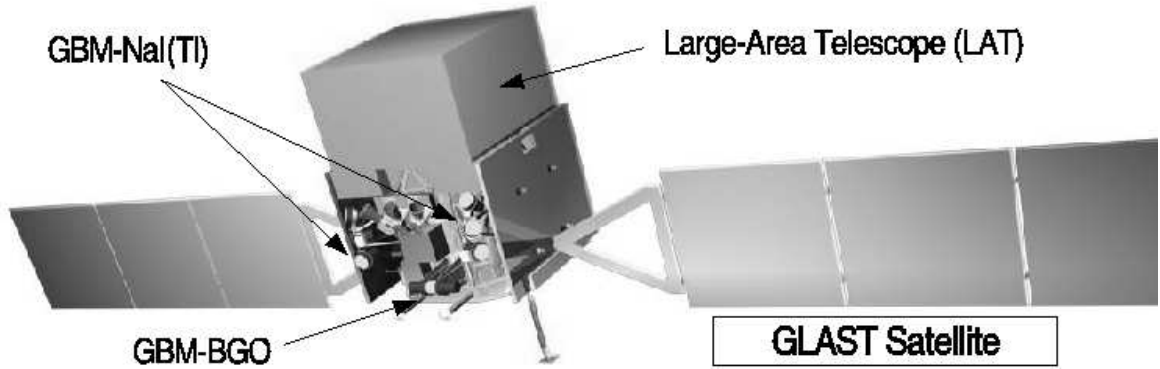}
   \end{tabular}
   \end{center}
   \caption[example] 
   { \label{fig:GBMview} 
A schematic view of the GLAST satellite with the GLAST burst monitor (GBM). The 12 NaI(Tl)-detetectors are 
mounted in 4 banks (the two banks on the back side are covered by the satellite), each equipped with 3 
NaI(Tl)-detectors. The two BGO-detectors are mounted on opposite sides of the satellite. In contrast to the 
NaI(Tl)-detectors each BGO-detector is viewed by two PMTs.}
   \end{figure} 

The main instrument on GLAST, the Large-Area Telescope (LAT), will itself detect bursts with high sensitivity, and 
locate them with a precision of about 10 arcmin. Within several seconds LAT's burst locations can be relayed to 
ground- and space-based observatories to search for afterglow emission. The expected LAT burst trigger rate is 
between 50 and 100 bursts/year. 

There are important limitations to the effectiveness of the LAT as a burst detector. High-energy measurements alone 
do not reveal the full physical picture. In the GRB energy spectra the most important characteristic currently known 
is a turnover or break (at break energy E$_{\rm{break}}$) between two parts of the spectra, each described by a power 
law, with different spectral indices, with $\alpha$ as the low-energy power-law index and $\beta$ as the high-energy 
power-law index \cite{1993ApJ...413..281B}; this break occurs in the energy range between 100 and 500 keV, well below 
the LAT threshold of about 20 MeV. Furthermore, $\gamma$-ray bursts have their maximal luminosity around the break 
energy. 
The scientific return of the GLAST mission in the case of $\gamma$-ray bursts will be increased substantially by 
having simultaneous knowledge of the burst emission from GeV down to a few keV. This will help to answer the open 
questions of the relation between the high- and low-energy emission and especially the question if the high-energy 
$\gamma$-rays are a part of the burst-emission process itself or a kind of afterglow. 

The GLAST Burst Monitor (GBM), which is able to cover the whole low-energy part of the $\gamma$-ray burst emission 
down to about 10 keV and simultaneously overlapping the lower part of the LAT energy range, will augment the 
capabilities of the LAT for $\gamma$-ray bursts. One of the important goals of the GBM is the continuation of the 
BATSE burst data base.
A significant concern for GLAST as a burst detector are the technical problems associated with triggering, 
rapid source location and dead time. With GBM as an auxiliary, autonomous instrument these limitations will be 
mitigated.
The development and fabrication of the NaI(Tl)- and BGO-detector modules and the power supplies (LVPS, HVPS)
is under the responsibility of  MPE/DLR\footnote{Allocation of funds by DLR: Deutsches Zentrum f\"ur Luft- und 
Raumfahrt (German Aerospace Center).}. The MSFC/UAH group is responsible for the development of the Data Processing 
Unit (DPU), hard- and software and the project management. 

\section{Instrument Description and Detector Design}

GBM consits of 12 thin NaI(Tl)-plates, which are sensitive in the
energy range between $\sim 10$~keV and $\sim 1$~MeV. Two additional BGO detetectors, which 
are able to detect gamma-rays in the energy range between 150 keV and
30 MeV, are responsible for the overlap in  energy with the LAT main instrument. 

The  12 NaI(Tl) and 2 BGO scintillation detectors are mounted on the spacecraft as shown in Figure~\ref{fig:GBMview}. 
The normals to the crystal discs of the 12 NaI(Tl) detectors are oriented in the following manner: six crystals in 
the equatorial plane (hexagonal), four crystals at $45^{\circ}$ (on a square) and two crystals at $20^{\circ}$ (on 
opposite sides). This arrangement results in a large field of view for the GBM of $> 8$ sr and gives the opportunity 
for locating the origin of the burst by comparing the count rates of different NaIs (same method as used by BATSE). 
The two BGO detectors will be mounted on opposite sides of the spacecraft, providing nearly a 4$\pi$ sr field of 
view.

   \begin{figure}
   \begin{center}
   \begin{tabular}{c}
   \includegraphics[width=0.45\linewidth]{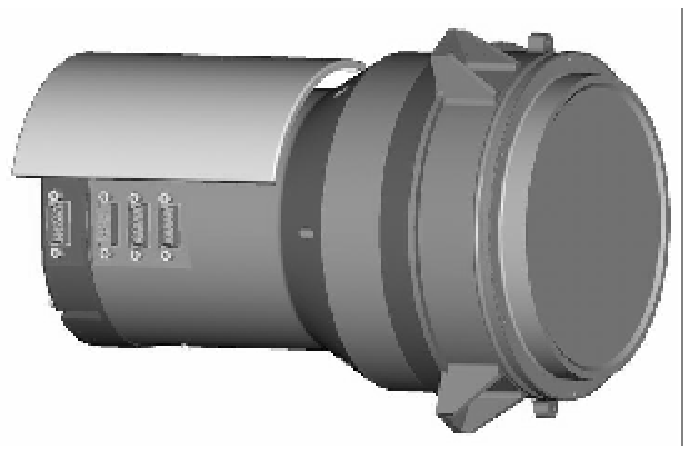}
   \includegraphics[width=0.45\linewidth]{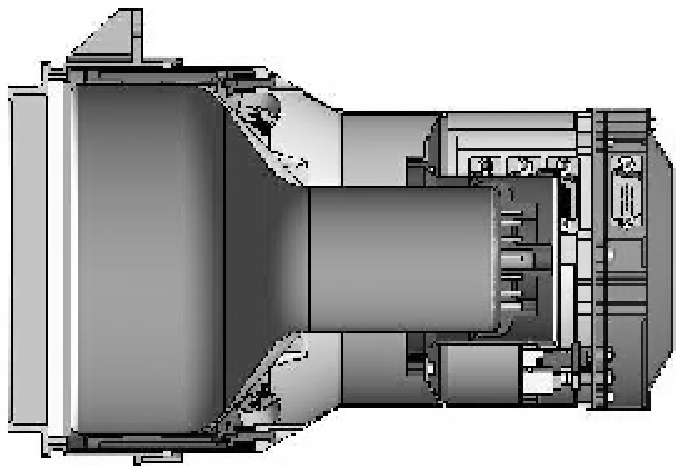}
   \end{tabular}
   \end{center}
   \caption[example] 
   {\label{fig:NaIdesign} 
Side view and cross-section of one of the 12 NaI(Tl)-detector modules.}
   \end{figure} 

Fig.\ \ref{fig:NaIdesign} presents the design of the NaI(Tl) detector unit. These detectors consist of circular 
crystal disks made from NaI(Tl) each disk having a diameter of 127 mm (5 inch) and a thickness of 12.7 mm (0.5 inch). 
For light tightness and for sealing the crystals against atmospheric moisture (NaI(Tl) is very hygroscopic) each 
crystal is packed light-tight and in a hermetically sealed Al-housing (with the exception of the glass window to 
which the Photomultiplier Tube (PMT) is attached). In order to allow the measurement of X-rays down to 5 keV, the 
radiation entrance window is made of a 0.2 mm thick Beryllium sheet. Opposite to the Be sheet a circular glass plate  
covers and seals the crystals. The inner sides of the packing material have a reflective white cover in order to 
increase the light output of the crystals.

For the use in the GBM detectors the phototube R877 of
Hamamatsu was selected. This tube was modified (R877RG-105) in order to fulfill the mechanical load-requirements for 
GBM. It is a head-on 5-inch diameter phototube made from borosilicate glass with a
bialkali (CsSb) photocathode. It is a tube with a box/grid dynode structure with 10 stages. This tube has a well 
defined
single-electron response and is therefore well-suited for detection down to very low energies.
It has a good stability with temperature, count rate and time. Each PMT is shielded against magnetic fields with  
0.125 mm thick Mumetal sheets, glued directly onto the glass tube. A second set of Mumetal sheets is attached to the 
PMT- and crystal-housing, which is further improving the magnetic shielding. The voltage divider, which is 
responsible for the supply of the dynode voltages, is mounted inside the PMT housing directly behind the feedthrough 
of the electrical leads. The collection efficiency of the photoelectrons is improved by using a divider ratio of 
2:2:1:1...1, which increases the voltage drop between the photocathode, grid and first dynode.  
The front end electronic box (FEE box), utilizing a charge-sensitive amplifier (CSA), pulse-shaping circuits $\&$ 
line drivers, is mounted on the rear side of each detector module.

   \begin{figure}[top]
   \begin{center}
   \begin{tabular}{c}
   \includegraphics[width=0.9\linewidth]{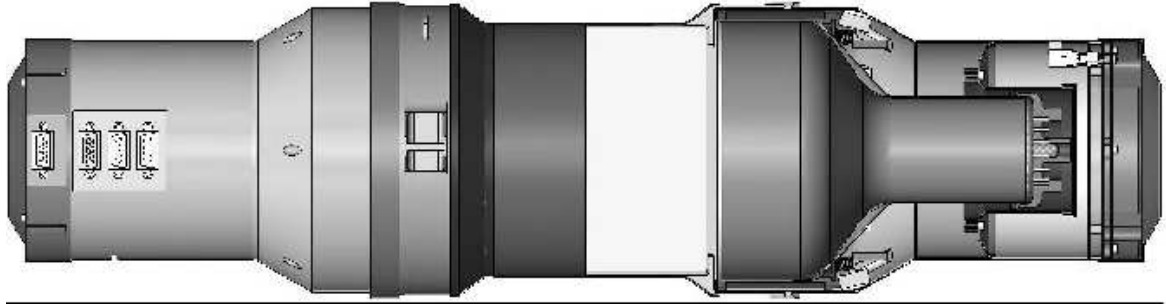}
   \end{tabular}
   \end{center}
   \caption[example] 
   {\label{fig:BGO-side-view} Side-view of the BGO detector unit. A bismuth germanate (BGO) scintillator-crystal, 
cylindrical in shape with a diameter of 127 mm (5 inch) and a length of 127 mm (5 inch), is viewed from both ends by 
a 5 inch PMT.}
   \end{figure} 

   \begin{figure}[b]
   \begin{center}
   \begin{tabular}{c}
   \includegraphics[width=0.33\linewidth]{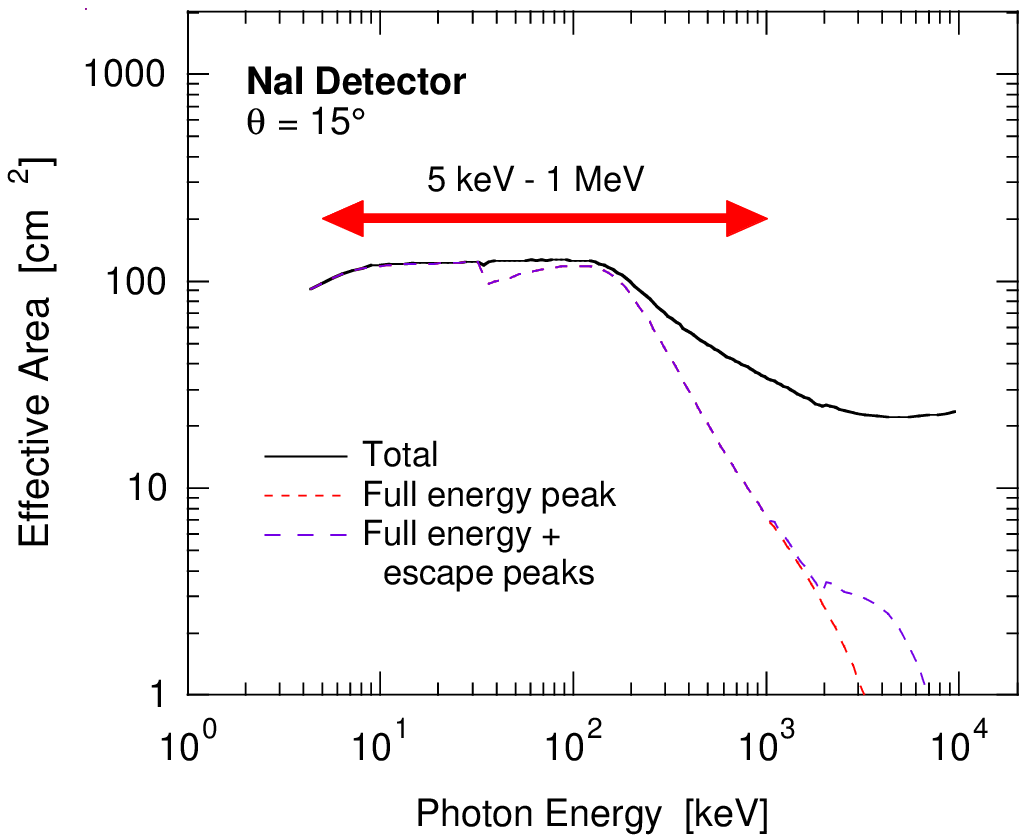}
   \includegraphics[width=0.33\linewidth]{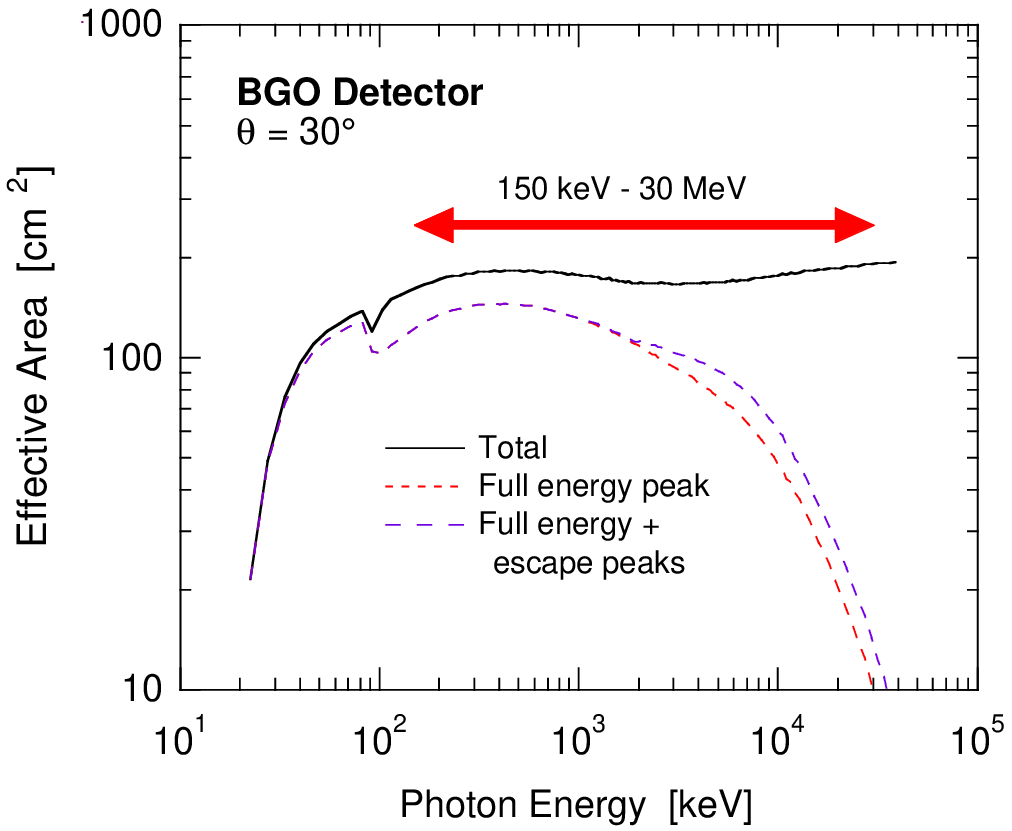}
    \includegraphics[width=0.3\linewidth]{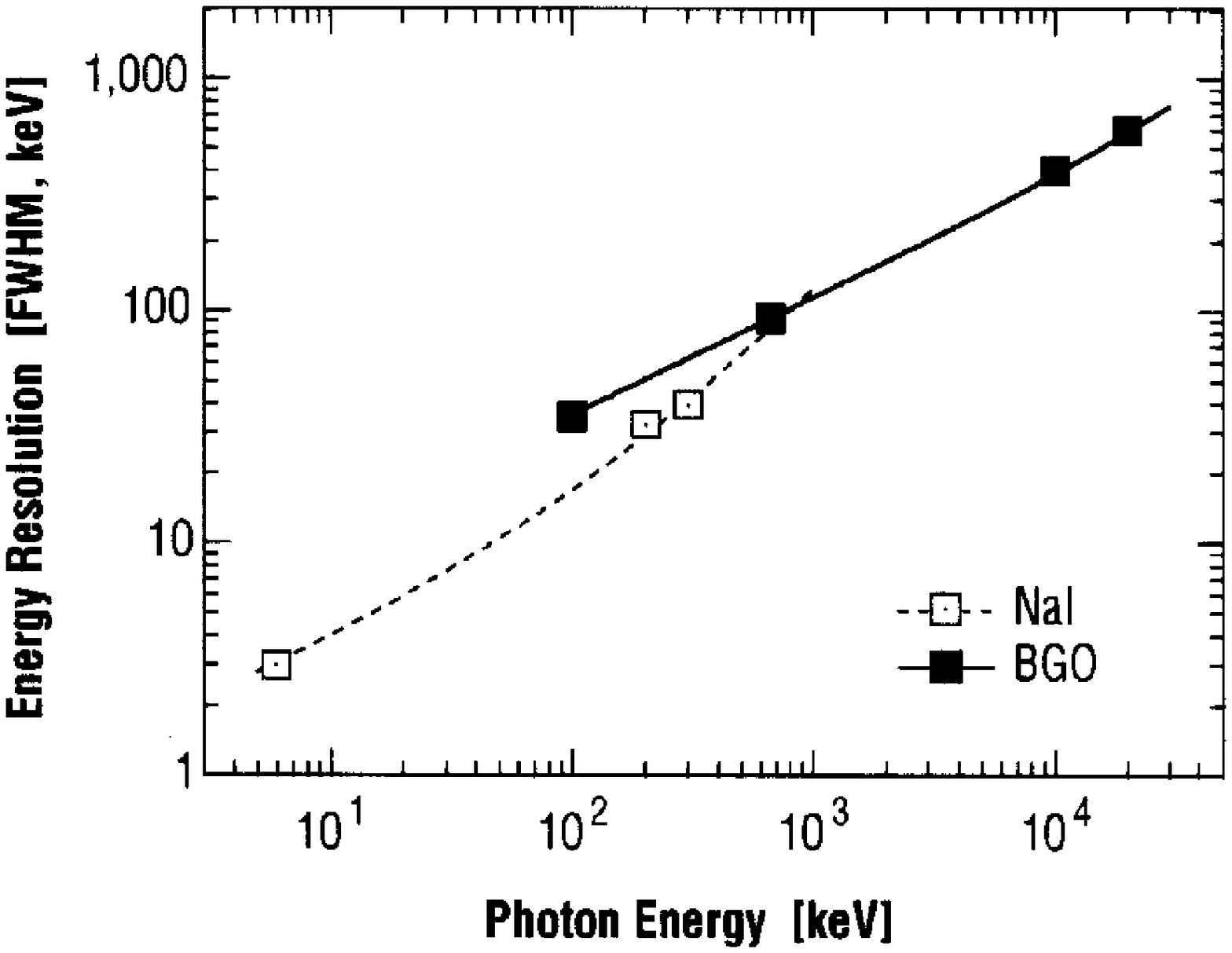}
   \end{tabular}
   \end{center}
   \caption[example] 
   { \label{fig:effA} 
Effective area of a NaI(Tl)- and BGO-detector in dependence on the photon energy, with $\Theta$ as angle of 
incidence. 
The double arrow shows the energy range of the NaI(Tl) (left graph) and BGO (middle graph) detectors. The right graph 
shows the energy resolution of the NaI(Tl)- and BGO-detectors in dependence on the photon energy.}
   \end{figure} 

The two bismuth germanate (BGO) scintillators, are cylindrical in shape, with a diameter of 127 mm (5 inch) and a 
length of 127 mm (5 inch). The BGO cover shown in Fig.\,\ref{fig:BGO-side-view} is made of CFRP (Carbon Fibre 
Reinforced Plastic) with interface parts made of titanium. The thermal expansion coefficient of these two materials 
match well to the expansion coefficient of BGO. The CFRP wrapping is providing the light tightness and improving the 
mechanical stability of the BGO unit. In contrary to NaI(Tl) BGO is not hygroscopic. The two circular side windows of 
the crystal are polished in mirror quality. The cylindrical surface is roughened in order to guarantee a diffuse 
reflection of the generated photons. The circular crystal windows are viewed by two PMTs (same type as used for the 
NaI(Tl) detectors), this guarantees a better light collection and a higher level of redundancy. The anode signals of 
both PMTs are summed at the input stage of the DPU. The expected response of the NaI(Tl)- and BGO-detectors, 
expressed in effective area and energy resolution, is summarized in Figure~ \ref{fig:effA}.

\begin{figure}
   \begin{center}
   \begin{tabular}{c}
   \includegraphics[width=0.85\linewidth]{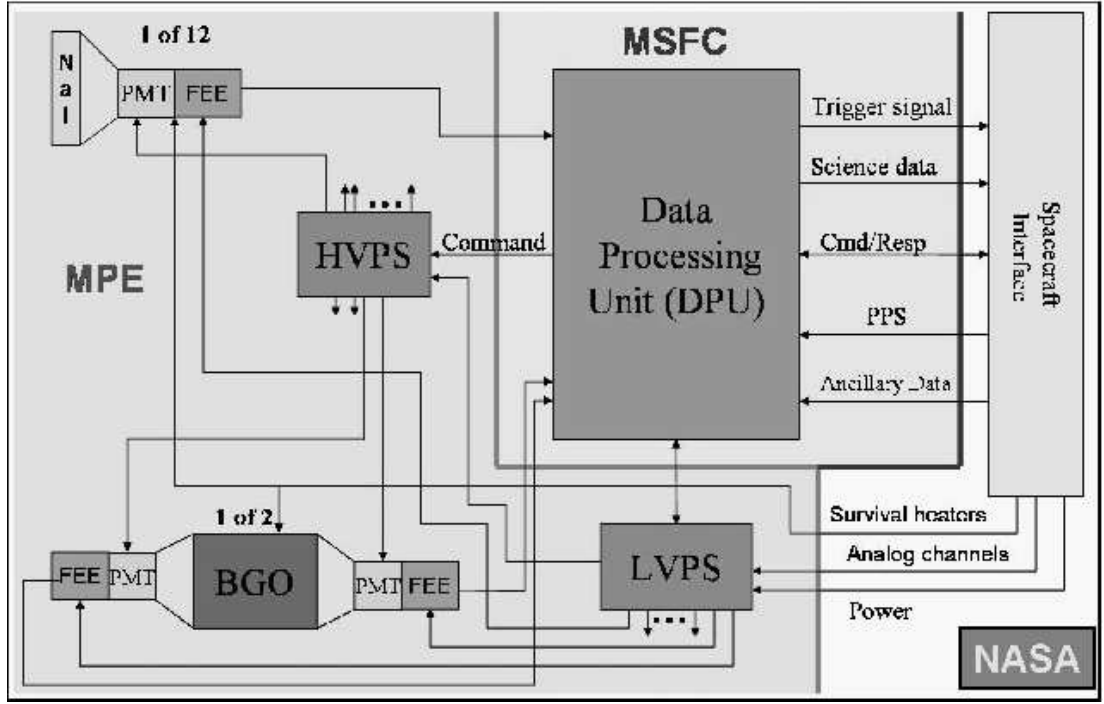}
   \end{tabular}
   \end{center}
   \caption[example] 
   {Block diagram of the main elements of the GBM subsystem. The responsibilities of MPE, MSFC and NASA are shown by 
different colours of the background. \label{fig:blockdiagram}}
   \end{figure}

The arrangement and electrical interconnections of the different main elements of the GBM subsystem is shown in 
Fig.\,\ref{fig:blockdiagram} as a block diagram. The 16 amplified PMT anode signals, 12 from the NaI(Tl) detectors 
and  $2 \times 2$ from the BGO detectors, are fed to the DPU for digitalization. The DPU software will search for a 
significant increase in the counting rates from the detectors. In case of a positive detection, a burst alert will be 
generated and transmitted to ground and to the main instrument. The DPU will calculate from the count rates of the 
NaI(Tl)-detectors a rough direction of the burst and will deliver this information to the main instrument as well. 
The central High Voltage Power Supply HVPS, providing the high voltages for the 16 PMTs, and Low Voltage Power Supply 
LVPS, providing the supply voltages for the 16 FEEs and the DPU, are built-in in the central Power Box (PB). The HV 
level applied to each PMT is commanded by the DPU. Gain changes of the PMTs, caused by temperature changes and aging, 
can be balanced out by the DPU automatic gain-control algorithm, monitoring the 511 keV background line. 
Likewise Fig.\,\ref{fig:blockdiagram}  illustrates the areas of responsibilities of MPE/DLR, MSFC/UAH and GSFC/NASA.

\section{Expected GBM Output/Results}
The tasks of the GBM can be split up into three main topics each yielding an output important for various
kinds of scientific analysis or for LAT support. 
The first output is certainly the GBM burst alert, which will be transmitted to LAT and ground.
Next is the burst position, which will be provided by the GBM over a wide field of view. The last 
and most important output is the low-energy context measurement of the burst light curve and burst spectrum.
\subsection{GBM burst trigger}
The trigger scheme for the GBM will be similar to that used with BATSE. The trigger requirement will be an excess in 
count rate above a threshold, specified in standard deviations above background, simultaneously for two of the 
NaI(Tl)-detector modules. The standard setting of the GBM threshold will be $4.5\,\sigma$ above background (energy 
interval: 50 keV to 300 keV, time interval for sensitivity calculations: 1.024 s). The requirement on the absolute 
on-board trigger sensitivity is  $<$ 1.0 photons cm$^{-2}$s$^{-1}$, with  $<$ 0.75 photons cm$^{-2}$s$^{-1}$ as a 
goal (BATSE at $5.5\,\sigma$ threshold: $\sim 0.2$ photons cm$^{-2}$s$^{-1}$). It is also planned to search on ground 
for fainter bursts using more sophisticated algorithms. One method is the summing of rates of closely pointing 
detectors and the inclusion of the BGO detector count rates.The current estimated sensitivity on ground is $\sim 
0.35$ photons cm$^{-2}$s$^{-1}$ ($5\,\sigma$ excess).

Based on the  ${\rm log}\,N - {\rm log}\,P$ burst intensity distribution determined by BATSE 
\cite{1999ApJS..122..465P} and considering the actual detector geometry, including the blockage by the LAT and 
spacecraft, the GBM will trigger on about 150-200 bursts per year. The estimated background level is based on BATSE 
rates and includes the effects of the variations due to altitude, latitude and longitude, and accounts for dead time 
due to transits through the South-Atlantic Anomaly (SAA). The simulation does not include the increased trigger rate 
one can expect by having additional triggering schemes that BATSE did not have, particularly the capability to 
trigger at lower energies.

\subsection{GBM burst localization}
The GBM determines locations of $\gamma$-ray burst by comparing count rates of
NaI(Tl)-detectors, which  are facing the sky in different directions. It is planned to
increase the location accuracy in three stages: on board, automatic on ground and on ground 
manually.
The burst location will be calculated on board in real time by the GBM-DPU, yielding an 
accuracy of about $< 15^{\circ}$ ($1~\sigma$ radius) within 1.8 s, which can be used as LAT trigger. If the burst 
occurred in the LAT field of view, 
data-reduction modes (reducing the LAT background by isolating the area of the GBM burst direction in the 
LAT dataspace) can be initiated in the LAT, which will increase the LAT sensitivity for weak bursts.
If the burst occurred  outside the LAT field of view (FoV), the spacecraft can be repointed to observe delayed
high-energy $\gamma$-ray emission. This is possible because the GBM FoV with $> 8$ sr is significantly 
larger  than the LAT FoV with approximately 3 sr.
After the transmission of the detector count rates to ground, the burst location can be computed with improved
accuracy of less than $5^{\circ}$ within 5 s. This will happen in near-real time (several seconds). This information
can be used for the search of afterglow emission at other wavelengths, as input for the Gamma-ray burst Coordinates 
Network (GCN) and as input for the Interplanetary Network (IPN).
The ground manual algorithms, which means a detailed analysis of the data with  human interaction, will yield an 
improved burst location $< 3^{\circ}$ after one day.  
%

   \begin{figure}
   \begin{center}
   \begin{tabular}{c}
   \includegraphics[width=0.6\linewidth]{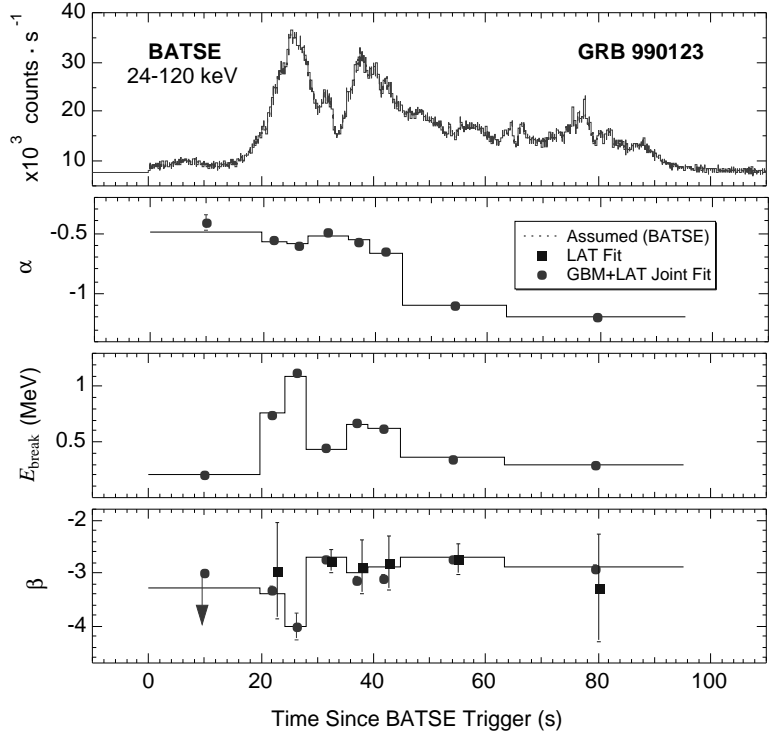}
   \end{tabular}
   \end{center}
   \caption[example] 
   {\label{fig:grb990123}
Time history of the simulated spectral parameters E$_{\rm{break}}$, $\alpha$ and $\beta$ 
using the data of GRB990123.}
   \end{figure} 

\subsection{GBM burst spectra and light curves}
The burst monitor will provide time-resolved spectra and energy-resolved lightcurves 
in the energy range between 10 keV and 30 MeV, overlapping with the LAT lowest energy range 
(low-energy threshold at $\sim$ 20 MeV).
In order to fulfill the scientific goals the burst monitor will have four main data types.
Two continuous data types are designed for burst analysis, for extremly long-lasting bursts,
search for non-triggered events and for the detection of bright sources via the Earth-occultation technique.
The first continuous data type accumulates 128 energy channels with 8.2 s time resolution for each detector and the 
second continuous data type 8 energy channels every 0.256 s. In response to a burst trigger, the GBM will produce 
a third datatype with high temporal resolution (2 $\mu$s) and 128 channel spectral resolution. The fourth data type 
provides information on the burst 
location and spectral estimates determined on board.

\subsection{Expected GBM results}
\begin{figure}
   \begin{center}
   \begin{tabular}{c}
   \includegraphics[width=0.475\linewidth,bbllx=40,bblly=360,bburx=550,bbury=665,clip=]{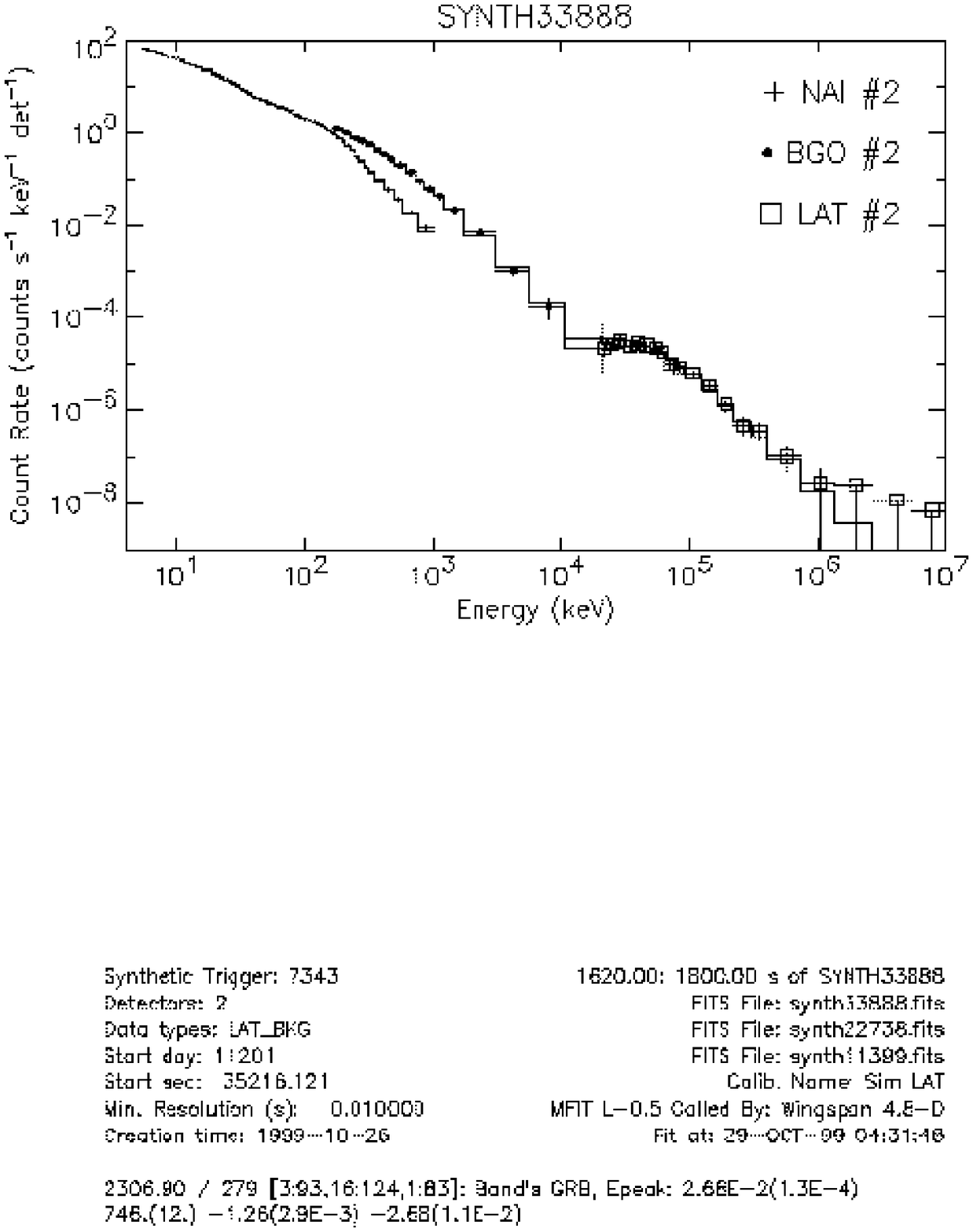}
   \includegraphics[width=0.475\linewidth,bbllx=35,bblly=360,bburx=550,bbury=675,clip=]{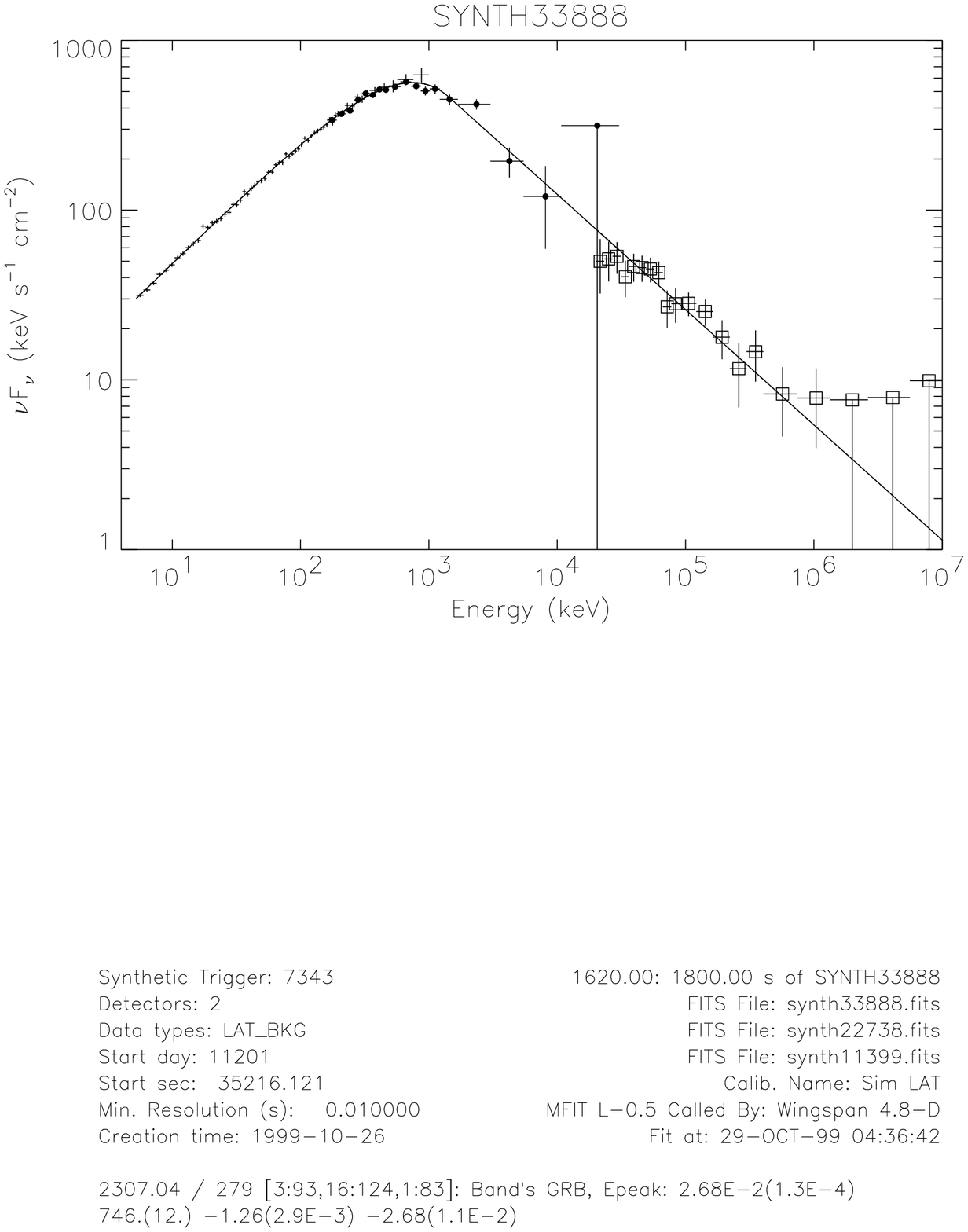}
   \end{tabular}
   \end{center}
   \caption[example] 
   {\label{fig:simflux} 
Simulated GBM-spectra using the data of GRB 940217 observed by BATSE, COMPTEL and EGRET.}
   \end{figure}

Preliminary simulations of GBM and LAT data were performed to estimate the GRB measurement capability of the combined 
instruments. Fig.~\ref{fig:grb990123} shows the results of one such simulation, wherein the time-resolved spectral 
parameters of GRB~990123 measured with BATSE were used as inputs.  This was one of the brightest bursts detected with 
BATSE, having a fluence of $3 \times 10^{-4}$~erg/ cm$^2$ ($> 20$~keV).  Looking at the results for the $\beta$ 
parameter one can see an excellent agreement, with small uncertainties, between the assumed BATSE values and the 
value derived from a common GBM/LAT fit.  In comparison, the LAT-only fit shows large uncertainties, and provides no 
information at all about the $\alpha$ and $E_{\rm{break}}$  parameters.  Fig.~\ref{fig:simflux} shows the estimated 
response of the GBM together with the LAT to the time-integrated energy spectrum of GRB~940217 measured by the three 
CGRO instruments BATSE, COMPTEL and EGRET.  The fluence of this burst was $7 \times 10^{-4}$~erg/ cm$^2$ ($> 
20$~keV).  The simulated spectral data cover 6 energy decades from $\sim 10$~keV up to $\sim 5$~GeV.  These examples 
indicate that the combination of GBM and LAT measurements promise to reveal the structure of prompt GRB spectra with 
unprecedented range and fidelity.
%

%
\section{Outlook}
The GBM together with the LAT will help to improve our understanding of the central engines and 
emission mechanisms of $\gamma$-ray bursts. Till now it is not understood how the high-energy 
emission observed by EGRET is related to the low-energy emission. Even the production mechanisms 
for these energetic $\gamma$-rays has not been explained by current models. How can they escape the 
source region without being absorbed via $\gamma$-$\gamma$ interaction with lower-energy photons?
The GBM will help to uncover how bursts, observed with LAT in the GeV regime, fit into the whole population.
It will also allow to link the LAT sample with the afterglow sample via the BATSE-like parameters. 
This is important, because the high-energy measurements alone do not allow a classification.   
It is one of the goals of the GBM team to produce a catalog of bursts that will include parameters such as 
location, duration, peak flux, and fluence, as well as spectral properties. These parameters 
will be defined as closely as possible to those in the BATSE catalog, so that the bursts observed by GLAST
can be related to the large sample of the BATSE catalog. The continuation of the BATSE database will 
be one of the benefits of the GBM. 
In contrast to the simple spectral shape of bursts the observed temporal behaviour is varying strongly.
Several common characteristic effects in the evolution of bursts have been observed, such as   
narrowing of pulses with increasing energy, a softening trend, and a hardness-intensity 
correlation. An interesting question for the GLAST mission will be how these temporal characteristics
will behave when the energy band is increased. 
The investigation of the temporal behaviour and distribution of the spectral parameters
(E$_{\rm{break}}$, $\alpha$, $\beta$ and the correlation between  E$_{\rm{break}}$ 
and the high-energy emission observed by the LAT) is one of the important tasks of the GBM.
There are several other questions which will hopefully be answered with the help of GBM:
In many spectra the power-law index  $\beta$  appears flatter than $-2$.
Such spectra cannot continue to infinitely high energies without steepening; 
an expected high-energy break may, in a few cases, be measured by the LAT alone. But in many cases
the constraints will be improved by fitting a wide-band spectrum, including the GBM spectra.
Another interesting question is whether there exists a significant population of hard-spectrum bursts 
which have been missed or poorly sampled by BATSE. GLAST can settle this question 
only by sufficient simultaneous coverage of the BATSE energy range.
The good temporal resolution will make GLAST an excellent detector for the 4$^{th}$ GRB Interplanetary Network (IPN), 
too.

\acknowledgments     
The Glast Burst Monitor  project is supported by the German "Ministerium f\"ur Bildung und Forschung" through 
DLR grant 50.QV.0301.


\bibliography{SPIEGBM1}   
\bibliographystyle{spiebib}   

\end{document}